# Evaluating Speech-in-Speech Perception via a Humanoid Robot

Luke Meyer[1,2,*], Gloria Araiza-Illan[1,2], Laura Rachman[1,2,3], Etienne Gaudrain[4], Deniz Başkent[1,2]

[1] Department of Otorhinolaryngology / Head and Neck Surgery, University Medical Center Groningen, University of Groningen, The Netherlands

[2] University of Groningen, University Medical Center Groningen, W.J. Kolff Institute for Biomedical Engineering and Materials Science, Groningen, the Netherlands

[3] Pento Audiology Centre, Zwolle, The Netherlands

[4] Lyon Neuroscience Research Center, CNRS UMR 5292, INSERM UMRS 1028, Université Claude Bernard Lyon 1, Université de Lyon, Lyon, France

**\* Correspondence:**
Luke Meyer
l.meyer@rug.nl

## Abstract

Underlying mechanisms of speech perception masked by background speakers, a common daily listening condition, are often investigated using various and lengthy psychophysical tests. The presence of a social agent, such as an interactive humanoid NAO robot, may help maintain engagement and attention. However, such robots potentially have limited sound quality or processing speed. As a first step towards the use of NAO in psychophysical testing of speech-in-speech perception, we compared normal-hearing young adults' performance when using the standard computer interface to that when using a NAO robot to introduce the test and present all corresponding stimuli. Target sentences were presented with colour and number keywords in the presence of competing masker speech at varying target-to-masker ratios. Sentences were produced by the same speaker, but voice differences between the target and masker were introduced using speech synthesis methods. To assess test performance, speech intelligibility and data collection duration were compared between the computer and NAO setups. Human-robot interaction was assessed using the Negative Attitude Towards Robot Scale (NARS) and quantification of behavioural cues (backchannels). Speech intelligibility results showed functional similarity between the computer and NAO setups. Data collection durations were longer when using NAO. NARS results showed participants had a more positive attitude toward robots prior to their interaction with NAO. The presence of more positive backchannels when using NAO suggest higher engagement with the robot in comparison to the computer. Overall, the study presents the potential of the NAO for presenting speech materials and collecting psychophysical measurements for speech-in-speech perception.

## 1    Introduction

Daily life often presents us with situations in which sounds with overlapping properties originating from different sources compete for our attention. Perception of speech in background noise requires segregating target speech and interfering masker signals. Further, in the case of competing background speech (speech masking), listeners need to suppress the information provided by the masking speech, oftentimes resulting in informational/perceptual masking (Carhart, Tillman, and Greetis 1969; Mattys, Brooks, and Cooke 2009; Pollack 2005). Speakers' voice characteristics facilitate segregating target speech from masking speech (Abercrombie 1982; Bregman 1990). Fundamental frequency (F0), related to the pitch of a speaker's voice (e.g., Fitch and Giedd 1999), and vocal-tract length (VTL), related to the size and height of a speaker (e.g., D. R. R. Smith and Patterson 2005), are two such speaker voice characteristics often used in differentiating voices and speakers (Gaudrain and Başkent 2018; Skuk and Schweinberger 2014). Normal-hearing listeners have been shown to be sensitive to small differences in F0 and VTL cues (Gaudrain and Başkent 2018; Koelewijn et al. 2021; El Boghdady, Gaudrain, and Başkent 2019; Nagels, Gaudrain, Vickers, Hendriks, et al. 2020), and can make effective use of these differences to differentiate between target and masker speech (Başkent and Gaudrain 2016; Darwin, Brungart, and Simpson 2003; Drullman and Bronkhorst 2004; El Boghdady, Gaudrain, and Başkent 2019; Nagels et al. 2021; Vestergaard, Fyson, and Patterson 2009). In contrast, hard-of-hearing individuals who hear via the electric stimulation of a cochlear implant (CI), a sensorineural prosthesis for the hearing-impaired, struggle in such situations, and show less sensitivity to F0 and VTL cues (El Boghdady, Gaudrain, and Başkent 2019). This challenge could be due to the inherent spectrotemporal degradation of electric hearing [see Başkent et al. (2016) for more information on the workings of CIs] and thus, difficulty in perceiving various speaker voice cues (El Boghdady, Gaudrain, and Başkent 2019; Gaudrain and Başkent 2018). Therefore, the investigation of these vocal cues through psychophysical testing, both in clinical and research settings, is important. On the other hand, evaluation of speech-in-speech perception requires the use of long and repetitive auditory psychophysical tests to ensure data reliability (Humble et al. 2022; Mühl et al. 2018; M. L. Smith et al. 2018). This can be a challenge for individuals being tested, especially for those with short or limited attention spans, such as young children (Bess et al. 2020; Hartley et al. 2000; Cervantes et al. 2023), or those with hearing loss, such as the elderly (Alhanbali et al. 2017). Therefore, any interface or setup that can improve engagement and focus may be helpful in collecting such data.

The use of a computer auditory psychophysics testing has led to methods that allow for better controlled experiments, more complex test designs, and the varying of more test parameters (Laneau et al. 2005). This has subsequently led to the use of desktop or laptop computers as typical test interfaces for auditory psychophysical tests (Marin-Campos et al. 2021; Zhao et al. 2022). When used as the test interfaces, the computer presents stimuli and collects responses. These capabilities have also expanded the potential use of computers for psychophysics testing outside of clinical or highly controlled environments (Gallun et al. 2018). Sometimes interfaces are modified to resemble a game-like format, especially for children (Harding et al. 2023; Moore 2008; Nagels et al. 2021; Kopelovich, Eisen, and Franck 2010). However, in a previous study by Looije et al. (2012), it was shown that during learning tasks, the use of a robot was better able to hold the attention of children in comparison to a computer interface. Furthermore, literature has shown that the physical presence of a social actor, both human and human-like, has a greater effect on engagement (Lee et al. 2006), in comparison to its virtual counterpart (Kidd and Breazeal 2004; Kontogiorgos, Pereira, and Gustafson 2021). This can be leveraged to motivate users to exert more effort during a given task (Bond 1982; Song et al. 2021). This was also reported by Marge et al. (2022), who comment that a robot can be advantageous in motivating and engaging users. Therefore, it could be that the inclusion of an interactive robot, such as the NAO humanoid robot, could be used to further retain one's attention, especially for psychophysical tests of speech-in-speech perception.

Over the last two decades humanoid robots have gained presence in a wide range of areas, including: high-risk environments (Kaneko et al. 2019; Sulistijono et al. 2010), entertainment (Fujita et al. 2003), home (Asfour et al. 2006), and healthcare (Choudhury et al. 2018; Saeedvand et al. 2019; Ting et al. 2014), to name only a few. Joseph et al. (2018) details more specifically how humanoid robots have been involved in healthcare applications, such as assisting tasks through social interactions (McGinn et al. 2014), telehealthcare (Douissard, Hagen, and Morel 2019), and nurse assistive tasks (Hu et al. 2011). The use of social robotics has steadily increased in recent years to the point where they are no longer only being used as research tools, but being implemented in day-to-day life (Henschel, Laban, and Cross 2021). The robots from Aldebaran Robotics (NAO and Pepper) are the two most frequently recurring robots in the field of social robotics. Moreover, the use of both the NAO and Pepper robots has been suggested in the literature as a facilitating interface in testing procedures for hearing research. Uluer et al. (2021), for example, have explored using a Pepper robot to increase motivation during auditory tests with CI children. The NAO has frequently been used in healthcare contexts, as shown in a scoping review by Dawe et al. (2019). Due to the robot's small size and its friendly and human-like appearance, the NAO has been used often in the investigation of child-robot interactions (Amirova et al. 2021). Polycarpou et al. (2016) used a NAO robot with seven CI children between the ages of 5 – 15 years to assess their speaking and listening skills through play. Although there have been other audiological studies utilising robot interactions, to the best of our knowledge, the evaluation or analysis of the human-robot interaction (HRI) has been limited and has predominantly focussed on task performance.

User engagement (Kont and Alimardani 2020) is one of the most frequently used metrics in human-robot interaction (HRI) analysis as it provides a measure of interaction quality, and thus one's perception toward an interface. One's own perception toward a robot is often performed using self-assessments, such as the Negative Attitude towards Robots Scale [NARS; (Nomura et al. 2004)]. The NARS is used to determine the attitudes one has towards communication with robots in daily-life and is divided into three components: subordinate scale 1 (S1), negative attitudes toward situations and interactions with robots; S2, negative attitudes toward social influence of robots; and S3, negative attitudes toward emotions in interactions with robots. In addition to self-assessments, much can be gleaned regarding the perception towards a robot as well as user engagement through the analysis of behavioural cues using video recordings. Verbal or gestural behavioural cues, known as backchannels and defined as cues directed back to a conversation initiator to convey understanding or comprehension, and a desire for the interaction to continue (Rich et al. 2010), have also been suggested as measures to evaluate user engagement (Türker et al. 2017).

In this study, we aim to expand the use of a NAO robot in psychophysical evaluations of speech-in-speech perception. Combined with its speech-based mode of communication, the NAO robot could be a relatively low-cost tool for auditory perception evaluation. In both research and clinical contexts, such an implementation could potentially provide participants with an interactive testing interface, possibly helping with engagement and enjoyment during experiments and diagnostic measurements (Henkemans et al. 2017). On the other hand, a number of factors related to the hardware and software of the robot could potentially affect auditory testing. For example, the internal soundcard and speaker combination may not be able to produce sound stimuli of sufficient quality for all psychophysical measurements (Okuno, Nakadai, and Kitano 2002), such as stimuli measured close to hearing thresholds. Non-experimental artefacts such as the noise of the fans or actuators in the robot could add unintentional background noise to the stimuli (Frid, Bresin, and Alexanderson 2018). Although the robot could potentially offer beneficial engagement during psychophysical tests, the different test setup with the NAO may impact the quality of the test results. Therefore, we first need to investigate how comparable the results are when conducting a psychophysics test using a robot to those when using the standard computer setup, while also evaluating the engagement factor via HRI analysis.

## 2 Methods

The present experiment is part of a large project, Perception of Indexical Cues in Kids and Adults (PICKA), and expands on previous work conducted using the same NAO robot for other psychophysical tests (Meyer et al. In press). The purpose of the PICKA project is to investigate the perception of voice and speech in varying populations, such as normal-hearing and hard-of-hearing adults and children with varying degrees and types of hearing loss and hearing devices, and in varying languages, such as English, Dutch, and Turkish.

In the present study, the PICKA speech-in-speech perception test was used. The speech-in-speech perception test evaluates speech intelligibility of sentences presented in competing speech, using an adapted version of the coordinate response measure (CRM, Bolia et al. 2000; Brungart 2001; Hazan et al. 2009; Welch et al. 2015). The test was performed via the computer [identical to that reported in Nagels et al. (2021)] as well as with a NAO humanoid robot named "Sam", chosen to represent a gender-neutral name. The computer and Sam versions of the test differ slightly in their implementation, much of which was done intentionally. The implementation differences are further explained in the sections below.

To compare the test performance with the robot to both the standard computer setup and to previous relevant work, we have collected both auditory speech intelligibility scores and data collection duration. To quantify the human-robot interaction (HRI), we have collected data in the form of a questionnaire, the Negative Attitude Towards Robots Scale (NARS), a common HRI metric (Nomura et al. 2004), and behavioural cues exhibited during the experiment to explore engagement related factors.

### 2.1 Participants

Twenty-nine (aged 19 – 36; 23.46 ± 4.40 years) individuals took part in the study. Two participants did not meet the inclusion criteria for normal hearing, and therefore data for the speech-in-speech perception test was analysed from 27 participants (aged 19 – 36; 23.23 ± 4.43 years). However, all 29 participants were included in the analysis of the HRI as there was no inclusion criteria for this component of the study. Sample size was determined based on a rule of thumb for human-robot interaction studies in which it is recommended that a minimum of 25 participants are included per tested condition (Bartneck 2020), and an extra four participants to account for potential drop-outs. All participants reported English as either native or additional language and completed at least high school education. A pure-tone audiogram was conducted to confirm normal hearing (NH). Hearing thresholds > 20 dB HL (Hearing Level) at any of the audiometric octave frequencies (between 250 Hz and 8 kHz) qualified for exclusion. The study was conducted according to the guidelines of the Declaration of Helsinki, and the PICKA project protocol was approved by the Medical Ethical Committee (METc) at UMCG (METc 2018/427, ABR nr NL66549.042.18). Written informed consent was obtained prior to the start of the experiment. The participants were compensated €8/hr for their participation.

### 2.2 Stimuli for speech-in-speech test

The CRM sentence stimuli used were in English, introduced by Hazan et al. (2009), Messaoud-Galusi, Hazan, and Rosen (2011) and Welch et al.(2015), and similar in structure to the Dutch sentences used by Nagels et al. (2021). The 48 English sentences contained a carrier phrase with a call sign ("dog" or "cat"), one colour keyword (selected from six colours: red, green, pink, white, black, and blue, all monosyllabic), and one number keyword (selected from eight numbers between 1–9, excluding disyllabic seven); e.g., *Show the dog where the pink (*colour*) five (*number*) is*. The same 48 sentences were used to create all stimuli for the present test. Each of the stimuli sets (Dutch and English) of the PICKA test battery were generated by a female speaker with a reference F0 of 242 Hz.

Target and masker sentences were originally produced by the same speaker. Speech-in-speech conditions were implemented by combining target and masker speech with two manipulations: 1) the target-to-masker ratios (TMRs) were varied, and 2) the voice cues F0 and VTL of the masker speech varied to introduce a voice difference between the target and masker speech [see Nagels et al. (2021) and El Boghdady et al. (2019) for a detailed explanation on the influence of TMR and voice cues on speech-in-speech perception]. For TMRs, expressed in dB, three conditions were used (-6 dB, 0 dB, +6 dB). F0 and VTL voice cues were expressed in semitones (st.), an intuitive frequency increment unit used in music and expressed as $1/12^{th}$ of an octave. Four different voice conditions were used: 1) the same voice parameters as the target speech, but with resynthesis to account for synthesis artefacts ($\Delta$F0: 0 st., $\Delta$VTL: 0.0 st.); 2 and 3) a difference of either -12 st. in F0 or + 3.8 st in VTL ($\Delta$F0: -12 st., $\Delta$VTL: 0.0 st.; $\Delta$F0: 0 st,. $\Delta$VTL: +3.8 st.); 4) a difference of -12 st in F0 and +3.8 st. in VTL ($\Delta$F0: -12 st., $\Delta$VTL: +3.8 st.). This resulted in 12 experimental conditions (three TMRs x four voice conditions). An additional condition with no manipulations (no TMR, no voice condition) was included as a baseline condition for a check of the experiment paradigm, but not included in data analyses. Each condition was tested with 7 trials (i.e., 7 target sentences), resulting in a total of 84 experimental trials + 7 baseline trials = 91 trials in the experimental corpus, all tested within one block.

For familiarisation of the test, a small corpus of training stimuli was created with nine F0 and VTL combinations: $\Delta$F0 = -12 st., $\Delta$VTL = 0.0 st.; $\Delta$F0 = -12 st., $\Delta$VTL = +1.9 st.; $\Delta$F0 = -12 st., $\Delta$VTL = +3.8 st.; $\Delta$F0 = -6 st., $\Delta$VTL = 0.0 st.; $\Delta$F0 = -6 st., $\Delta$VTL = +1.9 st.; $\Delta$F0 = -6 st., $\Delta$VTL = +3.8 st.; $\Delta$F0 = 0 st., $\Delta$VTL = 0.0 st.; $\Delta$F0 = 0 st., $\Delta$VTL = +1.9 st.; $\Delta$F0 = 0 st., $\Delta$VTL = +3.8 st.. The first two trials had a TMR of 0 dB and the remaining trials a TMR of +6 dB. Of the nine training stimuli, four were randomly selected for the training phase of the test.

For each trial, a target sentence was randomly selected from the 48 sentences with the "dog" call sign, and the masker speech was prepared from 48 sentences with the "cat" call sign. For the masker speech, random sentences were selected while avoiding sentences with the same number and colour keywords as the target sentence. From these sentences, 150 – 300 ms segments were randomly selected, applying 50 ms raised cosine ramps to prevent spectral splatter, and concatenating these segments to produce the masker speech. The masker speech started 750 ms before the target sentence onset and continued for 250 ms after the target sentence offset.

### 2.3 Human-robot interaction evaluation

The HRI was evaluated via the NARS questionnaire and behavioural data captured in video recordings of the experiment. The NARS is presented as a five-point Likert scale (1: strongly disagree – 5: strongly agree), used to grade each item, and the higher the score, the more negative an attitude one has toward robots. Total scores for each of the NARS subscales are obtained by totalling the grades of each subscale (S1, S2, S3). Therefore, minimum and maximum scores are 6 and 30 for S1, 5 and 25 for S2, and 3 and 15 for S3. For the video recordings, we analysed behaviours that could be used to indicate engagement (backchannels). "Smiling" and "laughing" (Türker et al. 2017) are two behaviours which can be considered positive backchannels and therefore positive engagement. To characterise negative backchannels, "frowning", and "grimacing" were used as opposites to smiling and laughing.

## 2.4 Setup

As mentioned previously, the paradigm of the speech-in-speech perception test is based on the CRM, which has been used extensively in the literature (Hazan et al. 2009; Welch et al. 2015; Nagels et al. 2021; Semeraro et al. 2017). In the standard version of the test, to log responses, participants make use of a coloured and numbered matrix representing all possible response combinations (Figure 1). Although other tests of the PICKA battery have been modified to resemble game-like interfaces (Nagels, Gaudrain, Vickers, Hendriks, et al. 2020; Nagels, Gaudrain, Vickers, Matos Lopes, et al. 2020; Meyer et al. In press), the speech-in-speech perception test has not been similarly modified to remain consistent with literature and allow for comparison to previously reported data.

### 2.4.1 Computer setup

The speech-in-speech perception test was run using MATLAB 2019b (MATLAB 2019) on an HP Notebook (Intel Core i5 $7^{th}$ gen) running Ubuntu 16.04. The user interface with the standard numbered matrix (Figure 1) was used, similar to Nagels et al. (2021). There are two deviations from the aforementioned study: English vs Dutch stimuli, and use of high-quality headphones vs internal soundcard and stereo speakers. We made use of the computer's loudspeakers in this study to present a more comparable test setup with the NAO, on which there is no audio connection for headphones.

[FIGURE 1 ABOUT HERE]

### 2.4.2 Robot setup

A NAO V5 H25 humanoid robot developed by Aldebaran Robotics (Sam) was used as an auditory interface to introduce the speech-in-speech perception test and present all corresponding stimuli. The PICKA Matlab scripts were rewritten into Python, which allowed all tests and stimuli to be stored and run directly on Sam. Housed in Sam is an Atom Z530 1.6 GHz CPU processor, 1 Gb RAM, and a total of 11 tactile sensors (three on the head, three on each hand, one bumper on each foot), two cameras and four ultrasound sensors (Figure 2, panel A). The software locally installed on the NAO robot is the NAOqi OS, an operating system based on Gentoo Linux specifically created for NAO by the developers. A cross-platform NAOqi SDK (software development kit) framework is installed onto a computer, which can then be used to control and communicate with the robot. The NAO SDKs available are Python (Van Rossum and Drake 2009), C++ (Stroustrup 2000), and Java (Arnold, Gosling, and Holmes 2005). NAO has 25 degrees of freedom and is able to perform movements and actions resembling that of a human.

[FIGURE 2 ABOUT HERE]

To improve the useability of running the PICKA tests through Sam, a simple website was designed for the researcher conducting any of the PICKA tests and hosted on Sam. Through this website, displayed on a Samsung Galaxy Tablet A, relevant participant information (e.g., participant ID and language) could be entered and the relevant PICKA auditory test could be initiated (Figure 2, Panel B). Stimuli were presented through the onboard soundcard, and the internal stereo loudspeakers located in Sam's head. The same tablet depicted a scaled down (approximately by a factor of 1.8) version of the aforementioned standard computer matrix for participants to log their responses (Figure 1). Henceforth, the robot and tablet are referred to as the "robot setup" and "auditory interface" refers to the robot only, as the tablet is considered a response logging interface.

### 2.4.3 Auditory interface calibration

The computer and the NAO inherently differ in their abilities to reproduce sounds due to the different hardware. To measure the output of the speakers, a noise signal that was spectrally shaped to match the averaged spectrum of the test stimuli was used. On both the computer and Sam, the noise was presented and measurements were recorded in third-octaves using a Knowles Electronics Mannequin for Acoustic Research (KEMAR, GRAS, Holte, Denmark) head assembly and a Svantek sound level metre (Svan 979). Measurements were conducted in a sound treated room, identical to that used for experimentation. The KEMAR was placed approximately one metre away from the auditory interface, similar to how a participant would be seated during the experiment. Replicating the experimental setup, the sounds were played on both interfaces at the calibrated level of 65 dB SPL (Figure 3). To further compare these signals, the digitally extracted levels from the original noise signal used for calibration have also been included in Figure 3 (blue line) to depict its spectral shape.

[FIGURE 3 ABOUT HERE]

Figure 3 shows that both the computer and Sam have relatively low level outputs below 250 Hz, compared to the original sound. Furthermore, the computer shows lower levels than Sam at frequencies below 800 Hz. To maintain the overall level of 65 dB SPL, this lack of low frequencies is then compensated above 800 Hz in the computer. These level differences will affect the perceived loudness and timbre of the sounds, and could also potentially affect audibility of lower harmonics in the speech stimuli.

### 2.4.4 General Setup

Participants were seated at a desk with either the computer or Sam and the tablet placed in front of them on the desk in an unoccupied and quiet room. Participants were seated approximately one metre from the auditory interface; however, this varied as participants moved to interact with Sam or the computer. The unused setup was removed from the desk and placed outside the participants' line of sight. To capture the behavioural HRI data, two video cameras were placed to the side and in front of the participant to capture their body positioning and facial expressions, respectively.

### 2.5 Procedure

Prior to their experimental session, participants were requested to complete the NARS questionnaire online. The order of the setups with which participants started the test was randomised. The speech-in-speech perception test consisted of two phases: a training phase and a data collection phase. The task was the same for both training and data collection. Participants were instructed that they would hear a coherent target sentence with the call sign "dog" that contained both a colour and a number (such as "*Show the dog where the red four is.*") in the presence of a speech masker to replicate a speech-in-speech listening scenario. Participants were also told that the speech masker might be louder, quieter, or have the same volume as the target, or be absent. Participants were instructed to log the heard colour and number combination on the provided colour-number matrix either by clicking with the connected mouse when using the computer or by touching the tablet screen.

Once the participant started the training phase and prior to the presentation of the first training trial, all stimuli for both the training and experimental corpora were processed with all TMR and voice conditions, and the splicing and resynthesis of speech maskers were randomised per participant. The training phase presented participants with four trials to familiarise themselves with the procedure of the test, but the participant responses were not taken into account for scoring purposes. Once confirmed by the researcher that the participant understood the test, the data collection phase started, consisting of a single block of all 91 trials (84 experimental + 7 baseline) with all sentences presented in a random order. Each logged response was then recorded as either correct or incorrect. Responses were only considered as correct when both the colour and number combination were correct. Participants performed the speech-in-speech perception test twice, once on each auditory interface with a break in-between, in a single session lasting approximately 40 minutes. Following the completion of the first iteration of the test on either the computer or Sam, participants were offered a break by the researcher before being seated again at the same desk with the next setup placed upon the desk.

When using the computer, participants were presented with the start screen of the test. Once "start" was clicked, the test immediately began with the training phase. Once completed, participants would again be presented with the start screen, which would initiate the data collection phase. No positive feedback was presented to participants; however, negative feedback was presented in the form of the correct colour-number pair briefly being outlined in green before continuing with the next trial. During the data collection phase, at predefined points, breaks would be offered to participants. A pop-up window would inform participants that they could take a break should they wish, and the test would resume when the pop-up window was dismissed.

When using Sam, the robot first introduced itself to the participant before explaining how the test would be carried out. Similar to the computer, first a training phase was presented to participants to familiarise themselves with the robot and the test procedure. Sam informed participants when the training phase was completed and waited for the participant to touch the top of its head to continue to the data collection phase. To maintain motivation and encouragement during the test, both positive (head nod) and negative (head shake) feedback were presented to participants throughout the training and data collection phases, as well as visual feedback to signal when a response could be logged (eyes turning green), and when the response was successfully logged (eyes return to default white). During the data collection phase, at the same predefined points as with the computer, a break was offered to participants. Sam would verbally ask the participant if they wanted to take a break, to which the participant could then verbally reply with either "yes" or "no". If the participant decided to take a break, Sam would ask a follow-up question if they would like to stand up and join in a stretch routine. Again, the participants could respond verbally with either "yes" or "no". If answered with "yes", Sam would stand and perform a short stretch routine. If answered with "no", Sam would stay in a seated position for 10 seconds before asking if the participant was ready to continue, again awaiting a verbal response. If "yes", Sam continued the experiment. If answered with "no", Sam would allow for another 10 second break before continuing the test. Once all trials were completed, Sam informed participants that they reached the end of the test and thanked them for their participation.

## 2.6 Data analysis

### 2.6.1 Test performance

Test performance was quantified by speech intelligibility scores (percentage correct) and data collection duration (minutes) with the computer and Sam setups. Intelligibility scores were calculated by averaging the recorded correct responses across all presented test trials per TMR and voice condition per participant. Data collection durations were calculated from when the first trial was presented until the response of the last trial was logged. Therefore, neither the interactions with Sam in the beginning and end of the test were taken into account, nor the duration of the training phases.

A classical repeated-measures ANOVA (RMANOVA) with three-repeated factors was performed for the intelligibility: the auditory interface with which the test was performed (computer or Sam), the four voice conditions applied to the masker voice ($\Delta$F0: 0 st., $\Delta$VTL: 0 st.; $\Delta$F0: -12 st., $\Delta$VTL: 0 st.; $\Delta$F0: 0 st., $\Delta$VTL: +3.8 st.; $\Delta$F0: -12 st., $\Delta$VTL: +3.8 st.), and the three TMR conditions (-6 dB, 0 dB, +6 dB), resulting in a 2x4x3 repeated-measures design. When RMANOVA tests violated sphericity, Greenhouse-Geisser corrections were applied ($p_{gg}$). Evaluation of data collection phase duration was performed using paired t-tests.

As the purpose of this study is to present a potential alternative auditory interface to the computer, we aim to look for evidence that both setups (using the computer and Sam) are comparable in their data collection. Therefore, for robustness, we also conducted a Bayesian RMANOVA using the same three-repeated factors as a conclusion of similarity cannot be obtained with classical (frequentist) inference. Bayesian inferential methods focus solely on the observed data, and not on hypothetical datasets as with classical methods. Therefore, they can provide an alternative interpretation of the data, the amount of evidence, based on the observed data, that can be attributed to the presence or absence of an effect [for more detailed explanations see (Wagenmakers et al. 2018)]. The output of Bayesian inferential methods is the Bayes factor (BF) and can be denoted in one of two ways: $BF_{01}$ where $0 < BF < 1$ shows increasing evidence for the null hypothesis as the BF approaches 0, and $BF > 1$ shows increasing evidence for the alternative hypothesis as the BF approaches infinity; and $BF_{10}$, which is the inverse of $BF_{01}$; i.e., $0 < BF < 1$ shows evidence for the alternative hypothesis, and $BF > 1$ shows evidence for the null hypothesis. The two notation methods can be used interchangeably for easier interpretation depending on the inference to be made. Since the intended focus of the inference of this study is evidence for the null hypothesis, the $BF_{10}$ notation is used. The degree of evidence is given by different thresholds of the BF: anecdotal, $0.33 < BF < 1$ or $1 < BF < 3$; medium, $0.1 < BF < 0.33$ or $3 < BF < 10$; strong, $0.03 < BF < 0.1$ or $10 < BF < 30$.

**2.6.2 Human-robot interaction**

Analysis of the NARS was performed using one sample t-tests were performed for each subscale to determine if the results were significantly different from the expected means (18, 15, and 9 for S1, S2, and S3, respectively), which would indicate neutrality toward interactions with robots, and thus an unbiased sample.

To analyse the behavioural data from the video recordings, two independent coders viewed the recordings and logged the frequency of displayed behaviours using the behavioural analysis software BORIS (Friard and Gamba 2016). Total duration of raw video footage was approximately 23 hours 57 minutes. To reduce the workload of coders, video recordings were post-processed and segments of different phases of the test were extracted. Segments were pseudo randomised and concatenated, resulting in approximately 8 hours 23 minutes of footage to be coded. Due to the repetitive nature of the test, these segments would provide "snapshots" during the different phases. Segments were created as follows: 35 seconds from the introduction when using Sam (introduction in its entirety); 30 seconds from the training phase for both the computer and Sam; two minutes from the beginning, one minute from the middle, and two minutes from the end of the data collection phase for both the computer and Sam; seven seconds from the break during the data collection phase in the case where the total duration was less than 10 seconds, or 45 seconds if the break was up to a minute. Engagement was assessed using the frequency of backchannels recorded by the two coders and were compared both within and between coders. Within coder comparisons were performed using Student t-tests. Reliability between coders was evaluated using intraclass correlation [ICC; (Bartko 1966)] based on the frequency of exhibited behaviours during each of the concatenated video segments. An ICC analysis is often used for ordinal, interval, or ratio data (Hallgren 2012). Because the frequency of behaviours is analysed per interval of the full video recording, as well as all subjects are observed by multiple coders, this makes an ICC appropriate.

## 3   Results

### 3.1 Test performance

#### 3.1.1 Speech intelligibility scores

The baseline speech intelligibility scores with no speech masker showed good consistency of the experimental paradigm: 99.0% on average when using the computer, and 99.5% on average when using Sam. Figure 4 shows the intelligibility scores per TMR and voice condition across all participants. Table 1 shows the results of both the classical and Bayesian RMANOVAs performed across both setups, three TMRs and four voice conditions. Results of the classical RMANOVA showed no significant difference between participants' intelligibility scores when using the computer or Sam [$F(1, 36) = 1.090$, $p = 0.306$, $n_p^2 = 0.040$], no significant interaction between the auditory interface and the TMR [$F_{gg}(1.490, 38.746) = 0.065$, $p_{gg} = 0.888$, $n_p^2 = 0.003$], no significant interaction between the auditory interface and the voice condition [$F_{gg}(2.353, 61.182) = 0.673$, $p = 0.537$, $n_p^2 = 0.025$], and no significant interaction between all three factors [$F(3.643, 94.730) = 0.587$, $p = 0.657$, $n_p^2 = 0.022$].

[TABLE 1 ABOUT HERE]

Bayesian RMANOVA showed moderate evidence that the auditory interface on which the test was performed did not affect the results of the speech-in-speech perception test ($BF_{10} = 0.185$), strong evidence of no interaction between the auditory interface and the TMR ($BF_{10} = 0.081$), strong evidence of no interaction between the auditory interface and the voice condition ($BF_{10} = 0.060$), and strong evidence of no interaction between all three factors ($BF_{10} = 0.039$).

[FIGURE 4 ABOUT HERE]

#### 3.1.2 Data collection duration

Figure 5 shows the duration of the speech-in-speech perception test when performed using each auditory interface, and in comparison, to previous data reported by Nagels et al. (2021). The average duration of the data collection phase was 9 ± 1 minute on the computer and 15 ± 5.1 minutes on Sam. However, we observed that three outlier participants took substantially longer to complete the data collection phase when using Sam. Removing these outliers resulted in an average duration of 13 ± 1 minute. The removal of the outliers showed that they had a significant effect on the total duration of the data collection phase [$t(45) = -12.22$, $p < 0.001$].

[FIGURE 5 ABOUT HERE]

### 3.1.3 Human-robot interaction

Results of the NARS questionnaire showed participants overall had a neutral attitude toward interactions with robots. Average scores for the subscales were 14.8 ± 3.74, 15.8 ± 2.17, and 8.5 ± 1.91 out of possible totals 30, 25, and 15 for S1, S2, and S3, respectively. One sample t-tests for each subscale showed only a statistically significant difference to the expected mean for S1 [$t(19) = -3.83$, $p < 0.01$], and nonsignificant differences for S2 and S3. The results are summarised in Table 2 below.

[TABLE 2 ABOUT HERE]

Behavioural coding results (Figure 6) showed on average (after pooling all backchannels) more frequent "frowning" when using the computer, although not statistically significant [$t(1.493) = 0.721$, $p > 0.05$], and significantly more frequent "smiling" when using Sam [$t(1) = -13$, $p < 0.05$]. "Grimacing" and "laughing" showed near identical frequencies between the two auditory interfaces. Intraclass correlation showed poor absolute agreement between coders for the behaviours "frowning" [$ICC(2,k) = 0.175$] and "laughing" [$ICC(2,k) = -0.375$], and high correlation for the behaviours "grimacing" [$ICC(2,k) = 0.671$] and "smiling" [$ICC(2,k) = 0.697$].

[FIGURE 6 ABOUT HERE]

## 4 Discussion

The aim of the present study was to evaluate Sam as an alternative auditory interface for the testing of speech-in-speech perception. To explore this, we compared the test performance data (both percent correct scores of intelligibility and data collection phase duration) obtained from normal-hearing young adults for the speech-in-speech perception test when using the proposed robot setup, to data when using the standard computer setup, as well as to previous studies using similar methods. Due to the inherent repetition of the speech-in-speech perception test, we propose Sam to offer an engaging experience for participants when conducting such a psychophysical test. Although there have been other studies in which psychophysical tests have been gamified to offer more engagement (Harding et al. 2023; Moore 2008; Nagels et al. 2021), there may be certain tests for which gamification may not be appropriate, either to be consistent with literature, or gamification may result in an overcomplication (e.g., Hanus and Fox 2015) of the test, having instead the opposite effect. In such cases, it may be beneficial to incorporate a social agent to facilitate engagement, not only due to its presence, but also playing an active role in the procedure. To explore this, we have also evaluated engagement with the two setups using an HRI questionnaire and analyses of behavioural data from video recordings.

### 4.1 Test performance

#### 4.1.1 Speech intelligibility scores

Results of the classical RMANOVA showed no significant difference between the percent correct scores obtained when using the computer or Sam. In addition, there was no significant interaction between the auditory interface and TMR, auditory interface and voice condition, or a combination of auditory interface, TMR and voice condition. Results of the Bayesian RMANOVA reflected the results of the classical RMANOVA, showing strong evidence in support of the two auditory interfaces being functionally identical. Visual inspection of Figure 4 also shows that the spread of the data between the TMRs and voice conditions are identical between the two auditory interfaces, and in comparison, to data reported by Nagels et al. (2021). It is also illustrated that most incorrect answers were given when the TMR was -6 dB, and a clear ceiling effect was observed at the TMR of +6 dB. The relatively higher percent correct scores for all conditions in the data reported by Nagels et al. (2021) could be due to several reasons. One possibility is that in their study the participants used high-quality headphones instead of the built-in loudspeakers of the computer. In addition, their stimuli were Dutch, whereas the stimuli presented to participants in the present study were English. Although Nagels et al. (2021) used Dutch stimuli, their population consisted of native Dutch-speaking participants. In the present study, participants reported English as either their native or an additional language. Therefore, the lower intelligibility scores seen in the present study in comparison to those reported by Nagels et al. (2021) may be due to a non-native effect. It is not expected that the structure of the CRM sentences would affect the intelligibility of the sentences since the paradigm of the sentence structure is intended to work across languages, as suggested by Brungart (2001). However, the English stimuli were presented by a British English speaker. This may have also affected the intelligibility of the target sentences in the presence of the masker sentences, especially in the -6 dB TMR condition, for the non-native English-speaking participants who may be more acquainted with US English, for example.

While we attempted to replicate the test procedure of Nagels et al. (2021) as closely as possible, as has been detailed above, there were some differences in the implementation of the test between the computer and Sam. Some of these implementation differences were necessary to perform a fairer comparison between the computer and Sam, but others were related to the interaction between the participant and Sam. These differences may have inadvertently introduced differences in the overall percent correct scores, resulting in the lower intelligibility scores.

In the present study, between the computer and Sam, stimuli, language, and target and masker speaker were kept consistent. Several factors were postulated to potentially limit the usability of the robot, such as the soundcard, speaker quality, processing speed, and non-experimental artefacts. An analysis of the speaker quality of the two auditory interfaces showed that there was a reduced quality of the computer in comparison to Sam, especially at lower frequency ranges. However, the consistent scores of the speech-in-speech perception test show that despite these limitations and the implementation differences between the computer and Sam, both setups were capable of presenting and collecting comparable test data. In addition, both the computer and Sam showed similar patterns in test results for the different TMR and voice conditions to those reported in literature. Therefore, the comparable results between the computer and Sam, and previously reported data, indicate that Sam can be used as an effective auditory interface for the speech-in-speech perception test with a normal-hearing population.

### 4.1.2 Data collection duration

The duration to complete the data collection phase of the speech-in-speech perception test was longer when using Sam in comparison to when using the computer; however, this increased duration did not seem to affect the performance of Sam's setup for collecting comparable intelligibility scores. The three outliers removed from the data collection duration were the first three participants with whom this test was conducted. During the experimental procedure with these participants, it was discovered that the pauses between stimuli were increasing. This was determined to be due to how response data was saved during the test; with each response given, the size of the save file increased, resulting in a longer duration to open and write to the file. Upon discovering this response saving issue, the test code was amended to save the results to a smaller file format during the test and subsequently saving the full results once the test was completed, thus rectifying the duration problem.

However, it can still be seen in Figure 5 that, even without the outliers, the duration of the data collection phase when using Sam was much longer than that when using the computer. We have considered several factors that could contribute to this difference. Potential delays due to online stimulus preparation were ruled out, since the stimulus corpus was created prior to the training phase. Further investigation into the data collection phase durations per participant showed that on average, there were six seconds between the logging of one response and the logging of the next response when using the computer. With Sam, however, this was on average nine seconds. Closer analysis of this three second difference showed that this occurs due to the feedback presented to participants following their response logging (head nod or head shake). Subsequent to the completion of data collection, separate measurements were taken by timing the duration of the head movements of Sam. On average, when a correct response was given, timings showed that it took 2.5 seconds for Sam to nod its head and then present the next stimulus. When an incorrect response was given, this time was on average 3.2 seconds for Sam to shake its head before presenting the next stimulus. Bootstrap simulations using the mean accuracy as the probability of a correct or incorrect response (and thus a head nod or head shake) for the various tested conditions showed that on average, the movement of Sam's head added 3.9 minutes ± 2 seconds over the 91 trials. No positive feedback and brief negative feedback (outlining of the correct response) was presented to participants when using the computer. The inclusion of positive and negative feedback when using Sam, although different to the computer implementation, was done to increase the social presence of the robot (Akalin, Kristoffersson, and Loutfi 2019).

## 4.2 Human-robot interaction

As mentioned previously, engagement during repetitive auditory tasks can be challenging, especially for certain populations, and to address this challenge we propose the use of a humanoid NAO robot. The use of such an interface for these tasks, at its core, relies on interactions, consisting of both social and physical components, between humans and the robot. The NARS questionnaire we used was developed as a measure of one's attitudes towards communication robots in daily-life (Nomura et al. 2004). The NARS is further broken down into three subscales to identify the attitudes of individuals toward social interactions with robots where the higher the score, the more an individual has negative attitudes towards those situations. The subscales are: S1, negative attitudes toward situations and interactions with robots; S2, negative attitudes toward social influence of robots; and S3, negative attitudes toward emotions in interactions with robots. Performing such a questionnaire prior to any interaction involving a robot allows it to be used as a cross-reference to explain any potential skewing of subsequently collected HRI data following the interaction. Results of the NARS subscales showed that only S1 was statistically different from the expected mean. The lower average S1 score indicates that participants had overall a more positive attitude towards situations of interactions with robots prior to their interaction with Sam. This is also reflected in the behavioural backchannels, coded from the video recordings. These showed more frequent smiling when using Sam in comparison to the computer, indicating both a state of comfort and engagement with Sam. This is contrasted by the more frequent frowning (although not significant, can be seen visually in Figure 6) when using the computer, which could indicate either a state of confusion (Rozin and Cohen 2003) or contemplation (Keltner and Cordaro 2017). Due to the nature of the speech-in-speech perception test and its fluctuating difficulty (especially when the TMR is -6 dB and where the target and masker speech did not differ in voice cues, the most difficult listening conditions tested), the more likely interpretation of the frowning is contemplation as participants focus harder in the more difficult voice conditions. Although this appears to be more frequent with the computer, this is not necessarily to say that the computer requires more focus. With both setups, this directed focus may subsequently lead to mental fatigue during the task (Boksem and Tops 2008). However, the results of the speech-in-speech perception test show that this increased directed focus does not affect the outcome of speech intelligibility between the computer and Sam.

## 4.3 General remarks

Our overall results show that the NAO robot shows promise to be used as an auditory interface for speech-in-speech testing. This finding is in line with and adds to our previous work (Meyer et al. In press), which evaluated the test performance from two other PICKA tests (voice cue sensitivity and voice gender categorization). Voice cue sensitivity test measures the smallest difference between two voice cues a listener can hear. The linguistic content seems to have little effect on the voice cue perception (Koelewijn et al. 2021), and the perceived voice could be biassed by the perceived gender of the robot (Seaborn et al. 2022). Speech-in-speech perception relies not only on processing voice and speech cues, but also on modulating attention and inhibition to separate target speech from masker speech, and further use of cognitive and linguistic mechanisms to decode the lexical content. It is not clear if a voice bias due to perceived robot gender would affect the speech intelligibility scores (Ellis et al. 1996). Despite such differing natures of these tests, our findings were consistent, and both showed comparable test performance with both setups.

## 4.4 Future directions

In comparing the test performance between the two setups, the only significant difference between the computer and Sam was the increased duration of the speech-in-speech perception test when using Sam. Although this is predominantly due to the presentation of positive and negative feedback to participants following the logged responses, we believe that it is an important component in establishing and maintaining the social presence of Sam. Therefore, instead of attempting to decrease the overall duration of the speech-in-speech perception test on Sam by removing the visual feedback, the social interaction with Sam could be improved. This way, we accept the longer duration with the inclusion of the feedback but provide the participant with a more natural interaction when performing the test. One such way this can be accomplished is by removing the use of the Samsung Galaxy tablet, which pulls the attention away from Sam with every response and replacing it with speech recognition on Sam. This would maintain the interaction with Sam both by not forcibly moving the participants' attention between Sam and the tablet, but also by engaging in more natural speech communication with Sam. The use of automatic speech recognition (ASR) for response logging has been explored in another study from our lab by (Araiza-Illan et al. in revision) with the use of Kaldi (Povey et al. 2011), an open-source speech recognition toolkit. The ASR was used to automatically score participant's spoken responses during a speech audiometry test. Their results show the robustness of the ASR when decoding speech from normal-hearing adults, offering a natural alternative for participants to give their responses throughout the test. Therefore, an ASR system, such as Kaldi, could be coupled with Sam, enhancing its social presence and overall interface functionality.

Literature has shown that the gamification of tests can also have beneficial effects on attention and engagement (Harding et al. 2023; Moore 2008; Kopelovich, Eisen, and Franck 2010). Although the speech-in-speech perception test has been suggested above to not be appropriate for gamification, it may indeed be interesting to explore how an intentional gamification of the test compares to the data collected here. This applies both to how speech intelligibility may be affected by such a gamification, but also how engagement may differ in comparison to Sam, especially after the implementation of speech recognition and removal of the tablet.

Both the present study and our previous work show the potential use of a NAO humanoid for speech-in-speech perception (present study) and voice manipulation perception (previous work) assessments by taking advantage of the robot's speech-related features. Furthermore, since current technical limitations are expected to be improved in the future, the proposed setup with the NAO provides exciting application possibilities in research and clinical applications.

## 5  Conflict of Interest

The authors declare that the research was conducted in the absence of any commercial or financial relationships that could be construed as a potential conflict of interest.

## 6  Funding

This study was supported by the VICI grant 918-17-603 from the Netherlands Organization for Scientific Research (NWO) and the Netherlands Organization for Health Research and Development (ZonMw). Further support was provided by the W.J. Kolff Institute for Biomedical Engineering and Material Sciences, University of Groningen, the Heinsius Houbolt Foundation and the Rosalind Franklin Fellowship from University Medical Center Groningen, University of Groningen.

## 7  Acknowledgments

We thank Josephine Marriage and Debi Vickers for sharing the English CRM stimuli, Paolo Toffanin, Iris van Bommel, Evelien Birza, Jacqueline Libert and Jop Luberti for their contribution to the development of the PICKA test battery, as well as Tord Helliesen and Conor Durkin for coding the video footage. The study was conducted in the framework of the LabEx CeLyA (ANR-10-LABX-0060/ANR-11-IDEX-0007) operated by the French ANR and is also part of the research program of the UMCG Otorhinolaryngology Department: Healthy Aging and Communication.

## 8  Data Availability Statement

The datasets generated (and subsequently analysed) for this study can be found in the DataverseNL repository (https://doi.org/10.34894/IAGXVF).

**List of Figures**

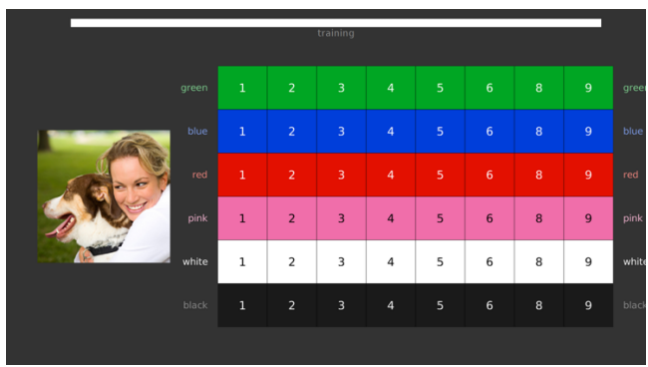

Figure 1: The standard computer user interface, showing the speech-in-speech perception coordinate response measure (CRM) test matrix as presented on the screen. Each item in the matrix represents a possible response option, corresponding to the target sentence. Bar at the top of the image depicts progress indicating how many stimuli are remaining in either the training or data collection phases. The matrix and image are published under the CC BY 4.0 licence (https://creativecommons.org/licenses/by/4.0/).

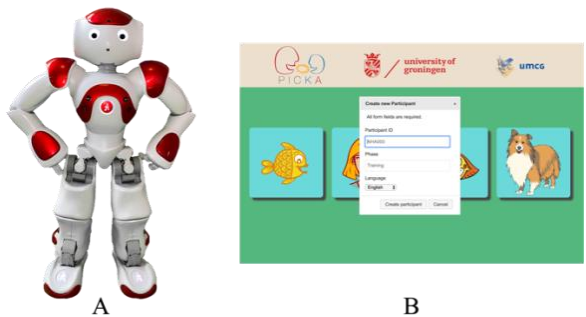

Figure 2: Panel A, The robot auditory interface, NAO V5 H25 humanoid robot from Aldebaran Robotics. Panel B, Webpage displayed on the Samsung Tablet to input participant details and begin one of the four PICKA psychophysics tests. Participant details included: participant ID, the phase of the test (either training or data collection), and the language of the test (either English or Dutch). Test buttons from left to right are for starting the different PICKA tests: voice cue sensitivity, voice gender categorization, voice emotion identification, and speech-in-speech perception (the focus of the present experiment), respectively. The cartoon illustrations were made by Jop Luberti for the purpose of the PICKA project. This image is published under the CC BY 4.0 licence (https://creativecommons.org/licenses/by/4.0/).

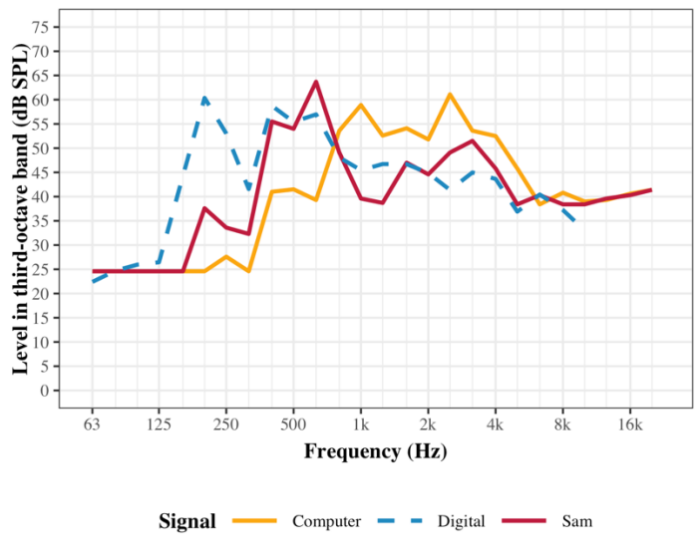

Figure 3: Comparison of auditory interface speaker comparison. Yellow and red lines show the levels of the noise signal when presented at the calibrated 65 dB SPL for the computer and Sam, respectively. Each point represents the total power within a third-octave band. The blue line is the digitally extracted levels from the noise signal and shifted to the ideal presentation of 65 dB SPL.

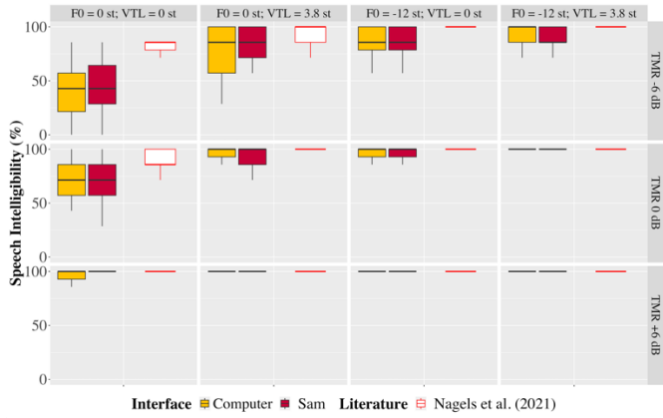

Figure 4: Boxplots depicting the range, quartiles, and median percent correct scores of the speech-in-speech perception test, shown for each talker-to-masker ratio (TMR, rows from top to bottom) and voice condition (columns from left to right) for the computer and Sam setups (yellow and red filled boxes, respectively), and in comparison to data reported by Nagels et al. (2021; empty boxes).

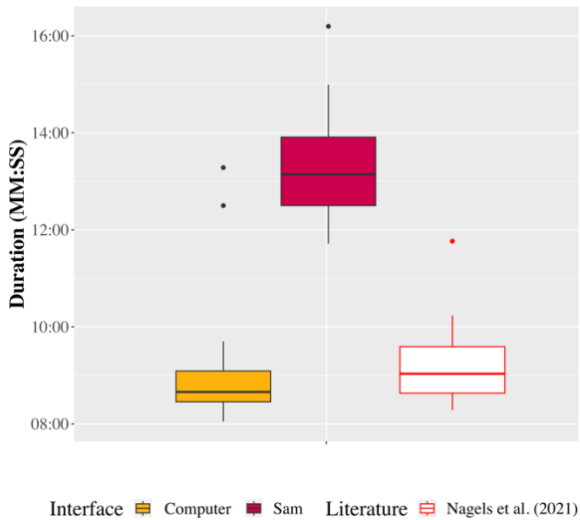

Figure 5: Duration to complete the data collection phase of the speech-in-speech perception test on the computer and Sam setups (following the removal of three outliers), and in comparison to data reported by Nagels et al. (2021).

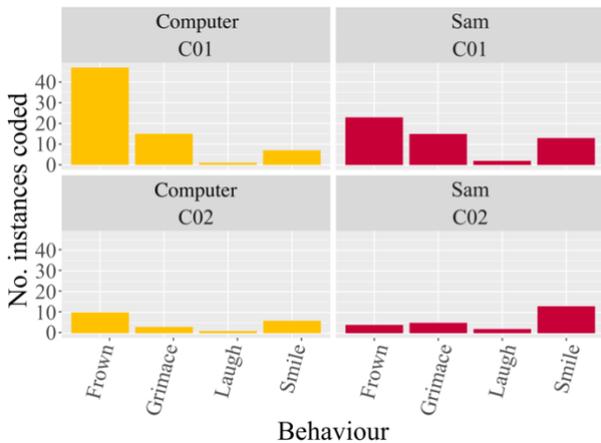

Figure 6: Coded backchannel frequencies by coder 1 (C01) and coder 2 (C02) when using both the computer and Sam.

**List of Tables**

Table 1: Results of the classical and Bayesian RMANOVAs.

|  | | | *Classical RMANOVA* | | | *Bayesian RMANOVA* |
|---|---|---|---|---|---|---|
|  | *Case* | *Sphericity Correction* | *F, p* | $\eta_p^2$ | | $B_{10}$ |
| *Main Factors* | Primary test interface | None | $F(1,36) = 1.090, p = 0.306$ | 0.04 | | 0.185 |
|  | TMR | Greenhouse-Geisser | $F(1.186, 30.826) = 147.980, p < 0.001$ | 0.851 | | 5.54E+18 |
|  | Condition | Greenhouse-Geisser | $F(1.982, 51.526) = 131.767, p < 0.001$ | 0.835 | | 3.50E+26 |
| *Interactions* | Primary test interface*TMR | Greenhouse-Geisser | $F(1.490, 38.746) = 0.065, p = 0.888$ | 0.003 | | 0.081 |
|  | Primary test interface *Condition | Greenhouse-Geisser | $F(2.353, 61.182) = 0.673, p = 0.537$ | 0.025 | | 0.06 |
|  | Condition*TMR | Greenhouse-Geisser | $F(4.002, 104.461) = 50.928, p < 0.001$ | 0.662 | | 2.99E+30 |
|  | Primary test interface*TMR*Condition | Greenhouse-Geisser | $F(3.643, 94.730) = 0.587, p = 0.657$ | 0.022 | | 0.039 |

Table 2: Results of the one sample t-tests comparing each of the NARS subscales to their respective expected means (indicating neutrality).

| *Questions* | *Subscale* | *Expected mean* | *Mean* | *SD* | *95% CI* | *t-score* | *p* |
|---|---|---|---|---|---|---|---|
| 1. I would feel uneasy if I was given a job where I had to use robots.<br>2. The word "robot" means nothing to me.<br>3. I would feel nervous operating a robot in front of other people.<br>4. I would hate the idea that robots or artificial intelligences were making judgements about things.<br>5. I would feel very nervous just standing in front of a robot.<br>6. I would feel paranoid talking with a robot. | S1 | 18 | 14.8 | 3.73 | [13.05, 16.55] | -3.83 | < 0.01 |
| 1. I would feel uneasy if robots really had emotions.<br>2. Something bad might happen if robots developed into living beings.<br>3. I feel that if I depend on robots too much, something bad might happen. | S2 | 15 | 15.8 | 2.17 | [14.79, 16.81] | -1.65 | N.S. |
| 4. *I am concerned that robots would be a bad influence on children.* | | | | | | | |
| 5. *I feel that in the future society will be dominated by robots.* | | | | | | | |
| 1. *I would feel relaxed talking with robots** | | | | | | | |
| 2. *If robots had emotions I would be able to make friends with them.** | S3 | 9 | 8.5 | 1.91 | [7.61, 9.39] | -1.17 | N.S. |
| 3. *I feel comforted being with robots that have emotions.** | | | | | | | |
| *\* inverse items* | | | | | | | |